# Abundant hydrocarbons in the disk around a very low-mass star

A. M. Arabhavi[1*], I. Kamp[1], Th. Henning[2], E. F. van Dishoeck[3,4], V. Christiaens[5,6], D. Gasman[6], A. Perrin[7], M. Güdel[8,2,9], B. Tabone[10], J. Kanwar[1,11,12], L. B. F. M. Waters[13,14,15], I. Pascucci[16], M. Samland[2], G. Perotti[2], G. Bettoni[4], S. L. Grant[4], P. O. Lagage[17], T. P. Ray[18], B. Vandenbussche[6], O. Absil[5], I. Argyriou[6], D. Barrado[19], A. Boccaletti[20], J. Bouwman[2], A. Caratti o Garatti[21,18], A. M. Glauser[9], F. Lahuis[22], M. Mueller[1], G. Olofsson[23], E. Pantin[17], S. Scheithauer[2], M. Morales-Calderón[19], R. Franceschi[2], H. Jang[13], N. Pawellek[8,24], D. Rodgers-Lee[18], J. Schreiber[2], K. Schwarz[2], M. Temmink[3], M. Vlasblom[3], G. Wright[25], L. Colina[26], G. Östlin[23]

[1]Kapteyn Astronomical Institute, University of Groningen, Groningen 9700 AV, The Netherlands.

[2]Max Planck Institute for Astronomy, Heidelberg 69117, Germany.

[3]Leiden Observatory, Leiden University, Leiden 2300 RA, The Netherlands.

[4]Max Planck Institut für extraterrestrische Physik, Garching bei München 85748, Germany.

[5]Space sciences, Technologies and Astrophysics Research Institute, Université de Liège, Liège 4000, Belgium.

[6]Institute of Astronomy, Katholieke Universiteit Leuven, Leuven 3001, Belgium.

[7]Laboratoire de Météorologie Dynamique, Centre National de la Recherche Scientifique, Palaiseau F-91120, France.

[8]Department of Astrophysics, University of Vienna, Vienna A-1180, Austria.

[9]Institute for Particle Physics and Astrophysics, Eidgenössische Technische Hochschule Zürich, Zürich 8093, Switzerland.

[10]Université Paris-Saclay, Centre national de la recherche scientifique, Institut d'Astrophysique Spatiale, Orsay 91405, France.

[11]Space Research Institute, Austrian Academy of Sciences, Graz A-8042, Austria.

[12]Fakultät für Mathematik, Physik und Geodäsie, Technische Universiteit Graz, Graz A-8010, Austria.

[13]Department of Astrophysics, Institute for Mathematics, Astrophysics and Particle Physics, Radboud University, Nijmegen 6500 GL, The Netherlands.

[14]SRON Netherlands Institute for Space Research, Leiden 2333 CA, The Netherlands.

[15]Free-Electron Lasers for Infrared eXperiments Laboratory, Institute for Molecules and Materials, Radboud University, Nijmegen 6525 ED, The Netherlands.

[16]Department of Planetary Sciences, University of Arizona, Tucson AZ 85721, USA.

[17]Université Paris-Saclay, Université Paris Cité, Commissariat à l'Énergie Atomique et aux Énergies Alternatives, Centre National de la Recherche Scientifique, Astrophysique Instrumentation et Modélisation, Gif-sur-Yvette F-91191, France.

[18]Dublin Institute for Advanced Studies, Dublin D02 XF86, Ireland.

[19]Centro de Astrobiología, Centro Superior de Investigaciones Científicas - Instituto Nacional de Técnica Aeroespacial, Villanueva de la Cañada 28692, Spain.

[20]Laboratoire d'Etudes Spatiales de d'Instrumentation en Astrophysique, Observatoire de Paris, Meudon 92195, France.

[21]Osservatorio Astronomico di Capodimonte, Istituto nazionale di astrofisica, Napoli 80131 Italy.

[22]SRON Netherlands Institute for Space Research, Groningen 9700 AV, The Netherlands.

[23]Department of Astronomy, Stockholm University, Stockholm 10691, Sweden.

[24]Konkoly Observatory, Research Centre for Astronomy and Earth Sciences, Budapest H-1121, Hungary.

[25]UK Astronomy Technology Centre, Royal Observatory Edinburgh, Edinburgh EH9 3HJ, UK.

[26]Centro de Astrobiología, Centro Superior de Investigaciones Científicas - Instituto Nacional de Técnica Aeroespacial, Torrejón de Ardoz E-28850, Spain.

*Corresponding author. Email: arabhavi@astro.rug.nl.

**Very low-mass stars (those <0.3 solar masses) host orbiting terrestrial planets more frequently than other types of stars, but the compositions of those planets are largely unknown. We use mid-infrared spectroscopy with the James Webb Space Telescope to investigate the chemical composition of the planet-forming disk around ISO-ChaI 147, a 0.11 solar-mass star. The inner disk has a carbon-rich chemistry: we identify emission from 13 carbon-bearing molecules including ethane and benzene. We derive large column densities of hydrocarbons indicating that we probe deep into the disk. The high carbon to oxygen ratio we infer indicates radial transport of material within the disk, which we predict would affect the bulk composition of any planets forming in the disk.**

Planets form in disks of gas and dust that orbit young stars. Observations have indicated that terrestrial planets form more efficiently than gas giants in the disks around very low-mass stars (VLMS), those with 0.08 to 0.3 solar masses ($M_\odot$) (*1-3*). Terrestrial planets orbiting VLMSs are therefore expected to be the most common type of planet around the most common type of star. The environment a planet forms in strongly influences its composition (*4-7*). Terrestrial planets are expected to form in the inner disk regions of VLMSs (<0.1 au from the host star, where au is the astronomical unit), which can be probed by observations at mid-infrared (mid-IR) wavelengths.

The chemical composition of the inner disk regions around higher mass stars have been studied previously (*8,9*). However, very few inner disk regions around VLMSs have been investigated (*10,11*). Infrared observations have shown that the ratio of molecular emission lines from $C_2H_2$ and HCN is <1 in disks around higher mass stars and >1 in disks around VLMSs suggesting a different evolution of gas in these disks (*10*). Weak or undetected lines of $H_2O$ indicate that those disks typically have a high carbon/oxygen (C/O) ratio (*11*) but there is one exception with strong $H_2O$ emission (*12*). A VLMS with a $C_2H_2/H_2O$ ratio several orders of magnitude larger than previously inferred has been found, indicating a C/O enhancement possibly by carbon-grain destruction thereby affecting the bulk composition of forming planets (*13*).

**Observations of ISO-ChaI 147**

ISO-ChaI 147 (also cataloged as 2MASS J11082650−7715550) is a VLMS located in the Chamaeleon I star-forming region. The median age of the objects in this star-forming region is 1 to 2 Myr and is located at a distance of ~191.4 parsec (1 parsec = 206,264.8 au) (*14*). ISO-ChaI 147 is a 0.11 $M_\odot$ and is classified as spectral type M5.5 (*15*). Previous observations detected $C_2H_2$ in the infrared emission (*10*). Dust emission at 887 μm was not detected from this target, indicating the mass of the millimeter-sized dust in the disk to be less than 0.72 earth-mass ($M_\oplus$) (*16*). No emission lines of CO or its isotopologues have been detected in the millimeter wavelengths either, setting an upper limit of 334 $M_\oplus$ for gas in the disk (*17*).

We observed ISO-ChaI 147 using Medium Resolution Spectrometer mode of the Mid-infrared Instrument (MIRI) (*18*) on the James Webb Space Telescope (JWST) (*19*). The observations were taken as part of the MIRI Mid-infrared Disks Survey (MINDS) (*20*,*21*); they cover the wavelength range 5 to 28 μm with a spectral resolving power ranging from ~3500 at 5 μm to ~1500 at 28 μm (*22*). We used a hybrid pipeline (*23*) to reduce the observed data to obtain a spectrum (*21*), which is presented in Fig. 1 after continuum subtraction.

**Molecules identified in the spectrum**

The mid-IR spectrum includes numerous emission lines, due to rotational or ro-vibrational transitions of molecules. We use the radiative transfer code PRODIMO (*24*) to calculate synthetic emission spectra for various molecules and gas properties (*21*). These were compared to the observed spectrum to determine which molecules are present. We identify 13 molecules in the spectrum, all containing carbon; their calculated contributions are shown in Figure 1. Nine (including isotopologues) of these are hydrocarbons (Table 1).

We detect ethane, which has previously been observed in the Solar System (fig. S5). We identify three *Q*-branch (emission with no change in rotational quantum number *J*, or $\Delta J = 0$) peaks of benzene and broader *P*- ($\Delta J = +1$) and *R*- ($\Delta J = -1$) branches (Fig. 1). To provide a broader context for these detections, the same benzene *Q*-branch lines have previously been observed in a disk around another VLMS, 2MASS J16053215−1933159 (hereafter J160532), as have $^{13}CCH_2$ and $C_4H_2$ (*13*). We also find $^{13}CO_2$ which has previously been observed in the disk around a higher mass star, GW Lup (*25*).

We do not identify any emission from silicate dust or polycyclic aromatic hydrocarbons (PAHs). There is a broad continuum emission in the 8.5 to 12 μm spectral range (fig. S1), but this cannot be explained by typical silicate dust (*21*). However, our synthetic emission spectrum of a high column density of $C_2H_4$ (Table 2) reproduces this wavelength region (Fig. 1). We do not identify any $H_2O$, OH, CO, atomic hydrogen (H I) or molecular hydrogen ($H_2$) emission lines. At wavelengths <6.6 μm, we identify stellar absorption features of CO and $H_2O$, indicating that the star dominates the spectrum at those wavelengths (fig. S6). We were unable to identify several emission lines in the spectrum (*21*).

**Estimated gas properties**

To determine the temperature, column density and emitting radius for each molecular species, we use the radiative transfer models to match one or two molecules at a time in the continuum subtracted spectrum. We also performed theoretical calculations of the incomplete molecular spectroscopic data for $C_3H_4$ and $C_6H_6$ (*21*). The resulting gas properties are listed in Table 2. Due to the overlap of various molecular emissions in the spectrum, our simple stepwise strategy to estimate the gas properties does not allow us to provide uncertainties following the usual $\chi^2$ framework (*21*).

We find large column densities ($>10^{22}$ cm$^{-2}$) of $CH_4$, $C_2H_4$, and $C_2H_2$, which produce the molecular continuum. We expect optically thick emission from these molecules to mask any more optically thin emission from other molecules in the same spectral and spatial regions. However, we identify emission lines of $C_6H_6$, $CO_2$, $^{13}CO_2$, $HC_3N$, and $C_2H_6$ superimposed on the molecular continua of $C_2H_2$ and $C_2H_4$. This implies that the emission from the less abundant molecules ($10^{17}$ to $10^{19}$ cm$^{-2}$) arises, at least partly, from regions where the $C_2H_2$ and $C_2H_4$ emission is also optically thin.

The estimated gas properties indicate that the gas temperatures are in the range of 200 K to 350 K, and the emission originates within 0.1 au of the star, which models have shown corresponds to the inner part of the disk (*26*). The temperatures and emitting radii are smaller than those previously determined for the disks around higher mass stars (*9*), consistent with the low luminosity of the VLMS. The similar temperatures of the abundant hydrocarbons suggest they arise from the same gas reservoir. These temperatures are cooler than was previously estimated for J160532, despite the similar stellar properties (see Supplementary Text).

The abundance of carbon-bearing molecules we identify, and the lack of oxygen-bearing molecules, indicate that the inner disk has a C/O ratio > 1. For comparison, the Milky Way's interstellar medium has an average C/O ratio of about 0.47 (*27*). We derive an upper limit on the water column density (*21*) but cannot calculate a corresponding upper limit for the C/O ratio without assuming co-spatial emission from all the detected molecules (which we consider unlikely) or upper limits on other oxygen-bearing molecules such as CO and $O_2$.

**Dust and molecular opacities**

Dust limits the observable gas in planet-forming disks. The absence of silicate features (at 10 μm and 18 μm) in the spectrum of ISO-ChaI 147 either indicates the dust grains in the inner disk are large (at least >5 μm in radius) (*28,29*) or that there is no silicate dust in the inner disk. In both cases, this implies a low dust opacity at mid-IR wavelengths. The molecular continua we identified indicate large column densities of gas. So the gas emission originates from deep in the disk (*26*), thus probing the terrestrial planet forming process.

**Origin of the hydrocarbons**

To investigate the origin of the observed hydrocarbon emission, we consider i) how the inner disk acquired a C/O ratio >1 and, ii) how the hydrocarbon molecules form once that C/O ratio has been reached.

*Increasing the C/O ratio*

Two scenarios have previously been proposed to produce a high C/O ratio in the inner disk: oxygen depletion and carbon enrichment (*7,13*). In the oxygen depletion scenario (Fig. 2A) pebble transport processes carry oxygen-rich ices to within the ice line (distance from the central star beyond which the volatile species are frozen on the dust surfaces), initially increasing the oxygen abundance in the inner disk through sublimation. However, this oxygen enriched gas is then rapidly accreted by the star, depleting the total oxygen content of the disk raising the C/O ratio of the remaining material. In this case, the composition of the solid dust grains remains unaffected. In the alternative carbon enrichment scenario (Fig. 2B), carbonaceous grains and PAHs are destroyed, releasing carbon into the gas phase in the inner disk, raising the C/O ratio. Potential destruction mechanisms include photochemical processes, high energy particles, combustion (reaction with oxygen), or thermal sublimation (see Supplementary Text). In this case, solid dust grains would be depleted in carbon and the oxygen abundance in the gas would be unaffected.

Dust evolution models predict that the dust in VLMS disks has short (~1 Myr) growth and dynamical timescales (*30*). The mass accretion rates for VLMSs indicate that the disk mass decreases by more than an order of magnitude in ~3 Myr (*5,31*). The sublimation of ice on the surface of large pebbles (coagulated dust) is slow enough to transport those ices within their respective ice lines (*32*). The ice lines in the VLMS disk are expected to be close (~0.1 au) to the host star, and the efficient ice transport produces a lower C/O in the inner disk than for the star itself (*7*). However, as the oxygen-rich inner disk gas is accreted onto the star, the C/O ratio then quickly (<2 Myr) switches to being higher in the inner disk than the host star (*7*). In this case the chemical state of the inner disk would reflect the dynamical history of the disk and the low abundance of pebbles would produce very little sub-millimeter continuum emission.

The low dust content in ISO-ChaI 147 should allow ultraviolet (UV) and X-ray photons from the central star to penetrate close to the mid-plane. Previous observations have shown the UV field is weak in this disk [accretion luminosity ~$4.47 \times 10^{-5}$ times the solar luminosity ($L_\odot$)] (*15*). By analyzing archival X-ray observations we derive, due to a non-detection, a $3\sigma$ upper limit on the X-ray luminosity $\lesssim 1.1 \times 10^{29}$ erg.s$^{-1}$ (*21*), and there is no detection of the [Ne II] 12.81 µm line (an indicator of gas illuminated by X-rays) (*33*). Hence, carbon enrichment by photodestruction is unlikely. This is consistent with the absence of PAH emission bands, which require UV photons for their excitation. High energy particles could be produced by flares of the young VLMS, then destroy carbon-rich grains, but the expected fluxes are unknown. If any of the potential carbon enrichment processes were efficient, they would also operate in at least some disks with inner dust cavities (known as transition disks) around young solar mass stars, but carbon-rich inner disks have not been observed in those objects (*9,34*). Our non-detection of OH and H$_2$O emission, and the estimated gas temperatures ($\lesssim 400$ K), make combustion or carbonaceous grain sublimation processes unlikely. We cannot exclude the possibility that the inner disk was hotter and more oxygen-rich in the past, converting most PAHs and carbonaceous solid dust into gas-phase CO and C$_n$H$_m$. This scenario was previously invoked to explain the carbon-rich disk around J160532 (*13*). However, the J160532 disk exhibits strong CO emission, unlike ISO-ChaI 147, so there could be different C/O enrichment mechanisms in these two sources (see Supplementary Text).

*Hydrocarbon chemistry in high C/O gas*

Regardless of how the gas in the inner disk reaches a high C/O ratio, hydrocarbons could either form in the inner disk gas or be transported inwards from colder outer regions of the disk. In high C/O environments, reactions can produce hydrocarbons on the surface of cold dust grains (see Supplementary Text), which could subsequently be transported to the inner disk. The current accretion rate of ISO-ChaI 147 [$7\times10^{-12}$ $M_\odot$yr$^{-1}$ (*15*)], indicates a currently small inward flux of material, though it could have been larger at earlier times. If large abundances of hydrocarbons were transferred from the outer disk, then hydrocarbon ices should be detectable in edge-on disks around VLMSs or in other high C/O astrophysical environments. In our solar system, though the comets are generally oxygen-rich, molecules such as $CH_4$, $C_2H_2$, $C_2H_6$, $C_6H_6$, HCN, $HC_3N$, and $CO_2$ have been detected in the comets 67P/Churyumov–Gerasimenko and C/2014 Q2 (Lovejoy) (*35,36*). The small molecules $C_2$ and $C_3$ are commonly observed in comets and have been suggested to form by the breakdown of $C_2H_6$, $C_3H_4$ or $C_4H_2$ (*37*). These comets probe the outer icy reservoir and thus support the picture of delivery of hydrocarbons to the inner disk.

The chemical timescales in the inner disk are expected to be short ($\lesssim$0.1 Myr) (*24*), so we expect the gas phase chemistry reaches a steady state. Gas phase astrochemical models with C/O>1 predict in-situ hydrocarbon formation with a similar molecular diversity as we observe in the spectrum of ISO-ChaI 147 (*38,39*) with large abundances of $CH_4$ and lower abundances of $C_2H_6$, consistent with our estimates (Table 2). However, the models also predict CO to be the most abundant carbon-bearing molecule in the gas phase in the inner disk where the mid-IR emission arises (*38*), which disagrees with our results. The weak emission lines from CO could be due to an emitting region smaller than those of hydrocarbons or the vibrational energy levels are not excited due to the low temperature and weak UV field. Models predict that a high C/O ratio in the disk by oxygen depletion would lead to a drop in the $CO_2$ flux by about an order more than the drop in $H_2O$ flux (*40*). This scenario alone does not explain our non-detection of $H_2O$ in the presence of strong $CO_2$ emission.

The most energetically stable hydrocarbons in pure C-H environments are referred to as magic compositions (*41*). All the hydrocarbons we observe in ISO-ChaI 147 have those magic compositions, indicating that they result from thermodynamic equilibrium in a high density and high-pressure environment. Assuming hydrostatic equilibrium and a carbon to hydrogen ratio of $3.55\times10^{-4}$ (*27*), we use the measured column densities to calculate maximum total gas pressures (at the bottom of the emitting gas column) of ~1 to 10 mbar. The pressure and temperatures inferred from these hydrocarbons indicate different formation conditions than those observed in atmospheres of Earth, Jupiter, and Titan (*42*). The disk around J160532 has different properties (lower column densities, higher temperatures) than that of ISO-ChaI 147, corresponding to roughly two orders of magnitude lower gas pressures, at which the gas-phase chemistry might not be in thermodynamic equilibrium. The features in the ISO-ChaI 147 spectrum that we were unable to identify (*21*) might be hydrocarbons with non-magic composition.

**Implications for planet formation**

Unlike gas giant planets whose atmospheric composition depends on both the solid and gas compositions of the disk, terrestrial planets form mainly from the solid materials. Simulations predict that the sizes and compositions of terrestrial planets are sensitive to the carbon abundance in disk's solid material (*43*).

Earth's low carbon abundance suggests removal of carbon from solids in the Solar System's protoplanetary disk (*44*), implying a higher gas-phase carbon abundance in the disk than the Sun's (*13*). The exchange of carbon between the solid and gas phases of the disk depends on whether the solid carbon is in the form of graphite, amorphous carbon, macromolecular carbon, or PAHs (*45,46*).

If the oxygen depletion scenario (Fig. 2A) is correct, it would not affect the carbon abundance in solids, but would produce planets with different compositions around VLMSs and higher mass stars, due to different dynamic timescales. The TRAPPIST-1 system consists of seven terrestrial planets orbiting within 0.1 au from the VLMS host (*47*). Models predict a wide range of bulk densities and high content of volatile elements for these planets (*48*,*49*). Any C/O enrichment processes that occurred during their formation would influence the planetary compositions. In case of carbon enrichment processes, the bulk composition of the planet would be carbon poor, like Earth, while oxygen depletion processes would not affect the carbon abundance.

**Tab. 1. Molecules detected in mid-infrared spectrum of ISO-ChaI 147.** Vibrational band identifications and notation follow (*50*); $\nu_x$:m-n indicates the transition of vibration mode 'x' from higher excitation level 'm' to a lower excitation level 'n'.

| Molecule | Vibrational band(s) | Wavelength [μm] |
|---|---|---|
| $CH_4$ | Degenerate deformation mode ($\nu_4$:1-0) | 7.65 |
| $C_2H_2$ | Asymmetric bending mode ($\nu_5$:1-0) | 13.69 |
| $^{13}CCH_2$ | Asymmetric bending mode ($\nu_5$:1-0) | 13.73 |
| $C_2H_4$ | $CH_2$ wagging mode ($\nu_7$:1-0) | 10.53 |
| $C_2H_6$ | $CH_3$ rocking mode ($\nu_9$:1-0) | 12.17 |
| $C_3H_4$ | CH bending mode ($\nu_9$:1-0) | 15.80 |
| $C_4H_2$ | CH bending mode ($\nu_8$:1-0) | 15.92 |
| $C_6H_6$ | Out-of-plane CH bending mode ($\nu_4$:1-0) | 14.85 |
| $CO_2$ | Fundamental bending mode ($\nu_2$:1-0)<br>Excited bending modes ($\nu_1\nu_2\nu_3$:100-010) | 15.00<br>13.88 & 16.18 |
| $^{13}CO_2$ | Fundamental bending mode ($\nu_2$:1-0)<br>Excited bending mode ($\nu_1\nu_2\nu_3$:100-010) | 15.41<br>16.20 |
| HCN | Fundamental bending mode ($\nu_2$:1-0)<br>Excited bending mode ($\nu_2$:2-1) | 14.00<br>14.30 |
| $HC_3N$ | HCC bending mode ($\nu_5$:1-0) | 15.08 |
| $CH_3$ | Out-of-plane bending mode ($\nu_2$:1-0) | 16.48 |

**Tab. 2. Estimated physical properties of the emitting molecules.** Temperatures (*T*), column densities (*N*), and emitting area equivalent radii (*R*) estimated from radiative transfer models (*21*) for the molecules detected in the ISO-ChaI 147 disk. For comparison, we also list those previously reported in the disk around J160532 (*13*). 'Detected' indicates species whose properties cannot be constrained in our fitting scheme (*21*). $\mathbb{N}$ is the number of emitting molecules (calculated as $N\pi R^2$). We find a systematic uncertainty of ±75 K and ±0.5 order of magnitude in temperatures and column densities respectively (*21*).

| Species | ISO-ChaI 147 disk | | | | J160532 disk | | | |
|---|---|---|---|---|---|---|---|---|
| | $T$ [K] | $N$ [cm$^{-2}$] | $R$ [au] | $\mathbb{N}$ [-] | $T$ [K] | $N$ [cm$^{-2}$] | $R$ [au] | $\mathbb{N}$ [-] |
| CH$_4$* | 300 | 1.00×10$^{23}$ | 0.048 | 1.62×10$^{47}$ | 400 | 1.50×10$^{17}$ | 0.070 | 5.20×10$^{41}$ |
| C$_2$H$_4$* | 350 | 3.16×10$^{22}$ | 0.031 | 1.45×10$^{46}$ | Not detected | | | |
| C$_2$H$_4$ | | 3.16×10$^{18}$ | 0.029 | | | | | |
| C$_2$H$_2$* | 325 | 2.15×10$^{22}$ | 0.031 | 2.13×10$^{46}$ | 525 | 2.40×10$^{20}$ | 0.033 | 1.84×10$^{44}$ |
| C$_2$H$_2$ | | 1.47×10$^{18}$ | 0.053 | | 400 | 2.50×10$^{17}$ | 0.070 | |
| C$_6$H$_6$ | 300 | 1.00×10$^{18}$ | 0.070 | 3.45×10$^{42}$ | 400 | 7.00×10$^{16}$ | 0.070 | 2.40×10$^{41}$ |
| C$_3$H$_4$ | 250 | 1.25×10$^{18}$ | 0.056 | 2.76×10$^{42}$ | Not detected | | | |
| C$_4$H$_2$ | 225 | 3.20×10$^{17}$ | 0.075 | 1.27×10$^{42}$ | 330 | 7.00×10$^{16}$ | 0.070 | 2.40×10$^{41}$ |
| CO$_2$ | 225 | 2.15×10$^{19}$ | 0.088 | 1.17×10$^{44}$ | 430 | 2.00×10$^{18}$ | 0.033 | 2.00×10$^{40}$ |
| C$_2$H$_6$ | Detected | | | | Not detected | | | |
| HCN | Detected | | | | 400 | <1.50×10$^{17}$ | 0.070 | <5.20×10$^{41}$ |
| HC$_3$N | Detected | | | | Not detected | | | |
| CH$_3$ | Detected | | | | Not detected | | | |
| H$_2$O† | 300 | <3.16×10$^{21}$ | 0.048 | <5.57×10$^{45}$ | 400 | <8.00×10$^{17}$ | 0.070 | <2.77×10$^{42}$ |
| CO | Not detected | | | | Detected | | | |
| H$_2$ | Not detected | | | | Detected | | | |

*These species form a pseudo-continuum in the ISO-ChaI 147 spectrum. †The upper limits are 1σ.

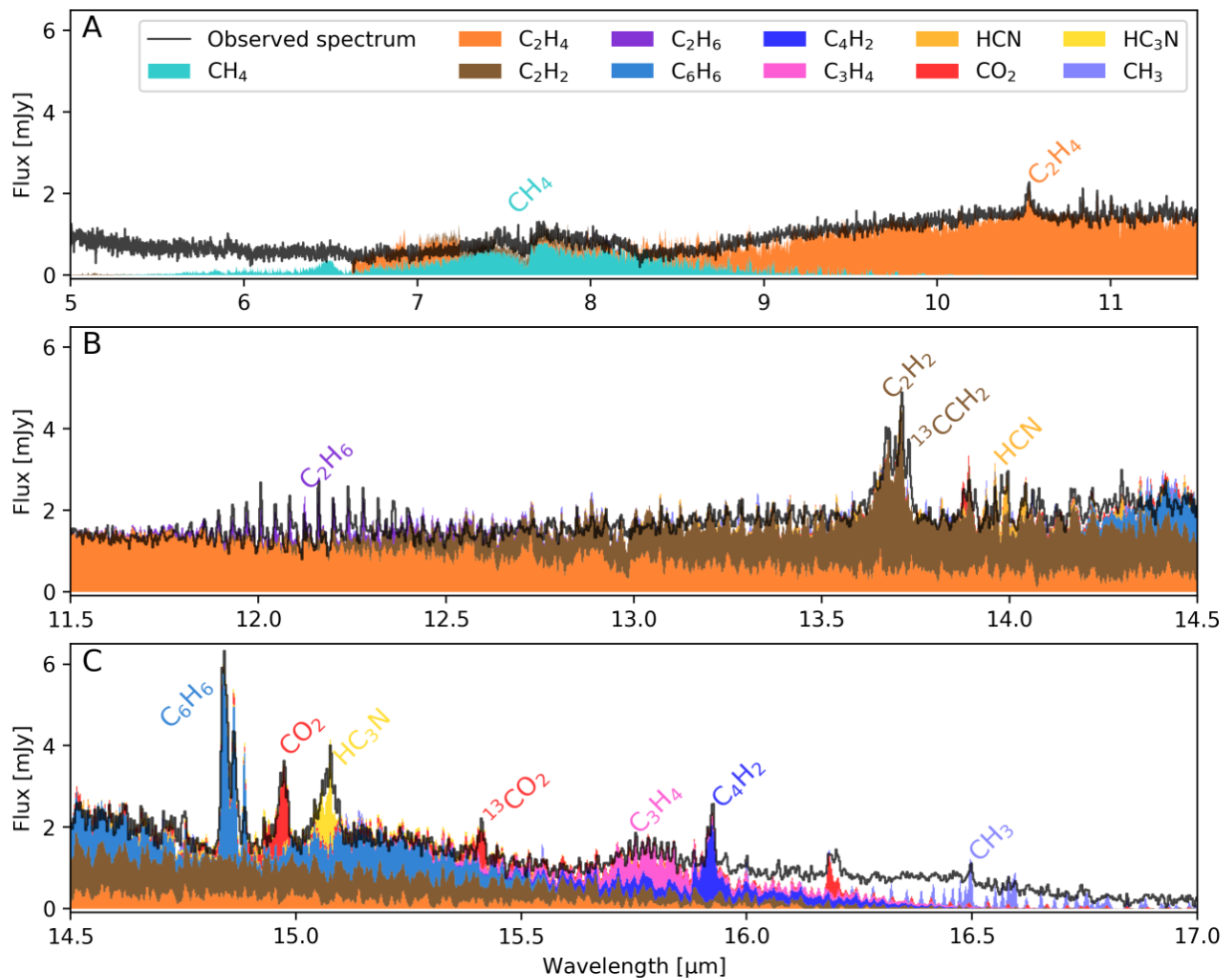

**Fig. 1. Mid-infrared spectrum of ISO-ChaI 147.** The black line is the continuum-subtracted MIRI spectrum (*21*). Colors indicate the modelled contributions to the spectrum of different molecules (labeled, see also the legend) which have been stacked. The estimated gas properties are listed in Tab. 2. Fig. S1 shows the same spectrum before continuum-subtraction (1 mJy = $10^{-29}$ Wm$^{-2}$Hz$^{-1}$).

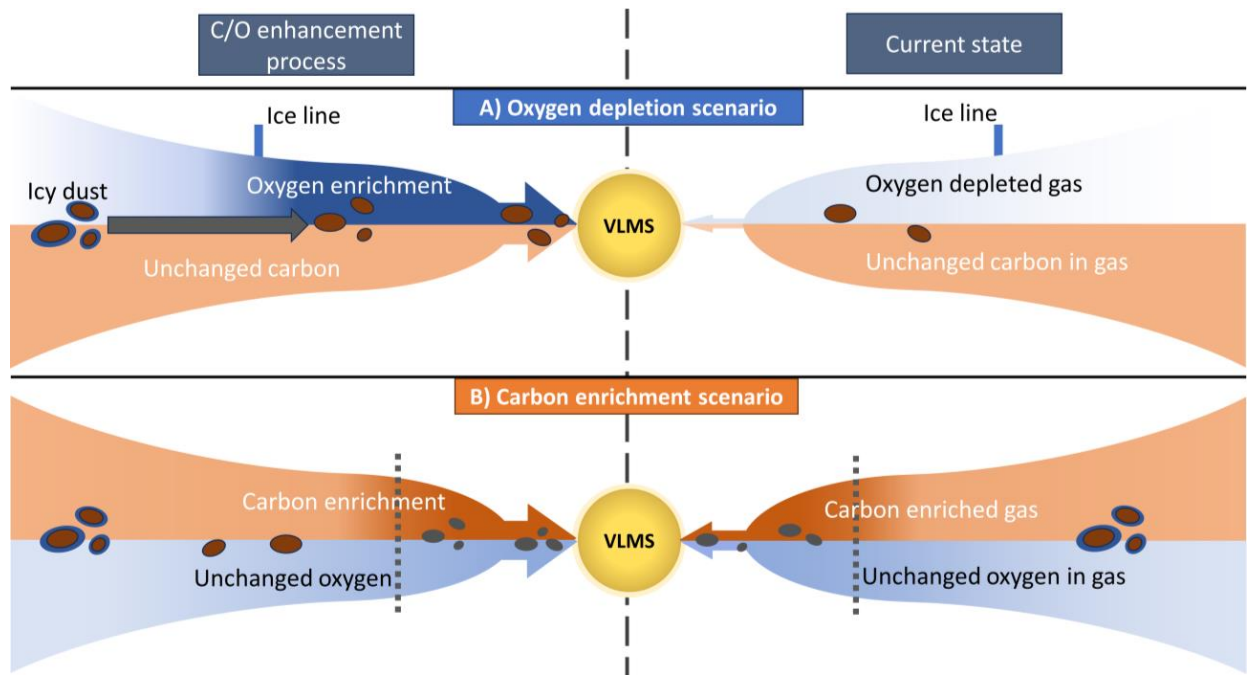

**Fig. 2. Two scenarios to produce a high C/O ratio in the inner disk.** Schematic cross-sections of the disk around a VLMS in the (**A**) oxygen depletion and (**B**) carbon enrichment scenarios. The left sides of each panel illustrate the C/O enrichment process, while the right sides show the resulting state of the disk, corresponding to when we observe it. Colored shading indicates elemental abundances of carbon (orange) and oxygen (blue). The two-color arrows pointing towards the VLMS indicate accretion of material onto the star. The grey arrow indicates transport processes. The region between the two vertical dotted lines indicates the region where the carbon in the solids is released into the gas. The brown, blue, and grey blobs indicate unaffected solids, ice covered solids and carbon depleted solids in the disk respectively.

**Acknowledgments:** Based on observations made with the NASA/ESA/CSA James Webb Space Telescope. The data were obtained from the Mikulski Archive for Space Telescopes at the Space Telescope Science Institute, which is operated by the Association of Universities for Research in Astronomy, Inc., under NASA contract NAS 5-03127 for JWST. We thank NASA, ESA, Belgian Science Policy Office (BELSPO), Centre Nationale d'Etudes Spatiales (CNES), Danish National Space Centre, Deutsches Zentrum fur Luft- und Raumfahrt (DLR), Enterprise Ireland; Ministerio De Economiá y Competividad, Netherlands Research School for Astronomy (NOVA), Netherlands Organisation for Scientific Research (NWO), Science and Technology Facilities Council, Swiss Space Office, Swedish National Space Agency (SNSA), UK Space Agency. Support from STScI is also appreciated. We thank Peter Hauschildt for providing the stellar model spectrum.

**Funding:**

IK, AMA, and EvD acknowledge support from grant TOP-1 614.001.751 from the Dutch Research Council (NWO).

VC and OA acknowledge funding from the Belgian F.R.S.-FNRS.

TH, RF, and KS acknowledge support from the European Research Council under the Horizon 2020 Framework Program via the ERC Advanced Grant Origins 83 24 28.

BT was funded by the Paris Region fellowship program, which is supported by the Ile-de-France Region and received funding under the Horizon 2020 innovation framework program and Marie Sklodowska-Curie grant agreement No. 945298.

DG was funded by the Research Foundation Flanders (grant number V435622N).

DG and IA thank the European Space Agency (ESA) and the Belgian Federal Science Policy Office (BELSPO) for funding as part of the PRODEX Programme.

IK and JK acknowledge funding from H2020-MSCA-ITN- 2019, grant no. 860470 (CHAMELEON).

EvD acknowledges support from the ERC grant 101019751 MOLDISK and the Danish National Research Foundation through the Center of Excellence "InterCat" (DNRF150).

TPR acknowledges support from ERC grant 743029 EASY.

DB was funded by Spanish MCIN/AEI/10.13039/501100011033 grants PID2019-107061GB-C61 and No. MDM-2017-0737.

ACG was funded by PRIN-INAF MAIN-STREAM 2017 and from PRIN-INAF 2019.

DRL acknowledges funding from Science Foundation Ireland (grant number 21/PATH-S/9339).

LC was funded by grant PIB2021-127718NB-I00, from the Spanish Ministry of Science and Innovation/State Agency of Research MCIN/AEI/10.13039/501100011033.

GÖ acknowledges support from the Swedish National Space Agency (SNSA).
**Author contributions:** TH and IK planned and co-led the MIRI observation, supported by MM, JB, FL, DB, LW, POL, TPR, BV, AB, and GW. AMA and IK analyzed the spectrum supported by TH, EvD, BT, LW, MM, RF, and JK. AMA, BT, SLG, and RF developed the methodology and software. VC, DG, GB, MS, GP, IA, JS, JB, MM, and SS developed the data reduction pipeline and AMA performed the data reduction, supported by VC, and DG. AMA, IK, TH, EvD, VC, IP, and MT wrote the manuscript. AP calculated the additional molecular data and wrote the related text. MG calculated the upper limit on X-ray luminosity. AMA, IK, TH, EvD, VC, DG, ACG, LW, AP, IP, MG, JK, BT, MS, GP, SLG, TPR, OA, RF, HJ, DRL, MMC, MT, MV, NP, GO, EP, LS, GÖ, and KS reviewed and edited the manuscript.

**Competing interests:** The authors declare that they have no competing interests.

**Data and materials availability:** The raw JWST data is available through the Mikulski Archive for Space Telescopes at https://mast.stsci.edu/portal/Mashup/Clients/Mast/Portal.html under program 1282. The reduced spectrum along with the continuum-subtracted spectrum and the synthetic molecular slab model fits are available as data S1. Our calculated molecular data for $C_3H_4$, and $C_6H_6$ are available as data S2, and data S3. Our modelling and analysis code (prodimopy version 2.2) is available at https://pypi.org/project/prodimopy/. The raw XMM-Newton data used in the paper is available through the XMM-Newton Science Archive at http://nxsa.esac.esa.int/nxsa-web/#search under observation ID 0203810201.

## Supplementary Materials

Materials and Methods

Supplementary Text

Figs. S1 to S6

Tables S1 to S4

Data S1 to S4

References (*51 – 92*)

# Supplementary Materials for

## Abundant hydrocarbons in the disk around a very low-mass star

A. M. Arabhavi[*], I. Kamp, Th. Henning, E. F. van Dishoeck, V. Christiaens, D. Gasman,
A. Perrin, M. Güdel, B. Tabone, J. Kanwar , L. B. F. M. Waters, I. Pascucci, M. Samland,
G. Perotti, G. Bettoni, S. L. Grant, P. O. Lagage, T. P. Ray, B. Vandenbussche, O. Absil,
I. Argyriou, D. Barrado, A. Boccaletti, J. Bouwman, A. Caratti o Garatti, A. M. Glauser,
F. Lahuis, M. Mueller, G. Olofsson, E. Pantin, S. Scheithauer, M. Morales-Calderón,
R. Franceschi, H. Jang, N. Pawellek, D. Rodgers-Lee, J. Schreiber, K. Schwarz, M. Temmink,
M. Vlasblom, G. Wright, L. Colina, G. Östlin
*Corresponding author: arabhavi@astro.rug.nl

**The PDF file includes:**

    Materials and Methods
    Supplementary Text
    Figs. S1 to S6
    Tables S1 to S4

**Other Supplementary Materials for this manuscript:**

    Data S1 to S4

**Materials and Methods**

Data reduction

ISO-ChaI 147 was observed with JWST-MIRI in medium resolution spectroscopy (MRS) mode (*19,51*) on 15 August 2022, from 23:57:04 Universal Time for a total observation time of 2.14 hr. This observation is part of the Cycle 1 Guaranteed Time Observation program (program ID: 1282, principal investigator: Thomas Henning, observation number: 51). Three grating settings short, medium, and long (referred to as A, B, and C respectively) were used in each of the four channels (channels 1, 2, 3, 4) following target acquisition, leading to 12 bands (1A, 1B, 1C, 2A, 2B, …, 4C). This provided the full coverage of the MIRI spectral window 4.9 μm to 28.1 μm. The FASTR1 read-out pattern was used with a point source four-point dither pattern. The total exposure time per sub-band was 1232 s.

We used a hybrid pipeline (*23*) to reduce our MIRI/MRS data. It relies on the standard JWST pipeline (v1.8.2) (*52*) using reference files context version 1061 from the JWST Calibration Reference Data System (*53*) combined with routines from the VIP package (*54,55*) to correct known data reduction issues not handled by the standard pipeline. Our hybrid pipeline is structured around three main stages that match the standard JWST pipeline (named Detector1, Spec2 and Spec3). After the first stage, we correct for stray light using the standard pipeline, and subtract the background using pair-wise dither subtraction. We then employ point-source optimized reference spectrophotometric and custom fringe flat files in Spec2 following previously described methods (*56*). The outlier detection step in the standard Spec3 was skipped, because it results in spurious artifacts for unresolved sources observed with the MRS due to the under-sampling of the point spread function (PSF). To compensate for this, we use VIP-based routines for bad pixel correction directly before and after Spec3. VIP-based routines are then also used for the identification of the star centroid in the spectral cubes produced by Spec3. The identified centroid locations are subsequently used for aperture photometry, where the spectrum is extracted by summing the signal in a 1.8 $\lambda/D$ aperture centered on the source (where $\lambda$ is the wavelength and $D$ is the diameter of the telescope, which is ~ 6.5 m for JWST). Aperture correction factors are applied to account for the flux loss (*18*). Finally, we compensated for the known spectral leak in band 3A (*56*). The source is faint at long wavelengths, leading to low signal in channel 4, so we limit our analysis to wavelengths shorter than ~18 μm. The reduced spectrum is shown in fig. S1. No spatially extended emission is found on scales larger than 0.24 arcsec (at 5 μm) - 0.86 arcsec (at 18 μm).

We used the spectrophotometric and fringe flat reference files that provide better de-fringing for point sources than the standard reference files (*23,56*). They avoid the need for the additional residual fringe step included in the standard pipeline which can change the shape of molecular features when they have similar frequencies as the expected fringes (*56*). Avoiding this step preserves the hydrocarbon features.

To remove the background, we choose to use a pair-wise dither subtraction. This method is more suited to fainter sources where the resulting PSF overlap is minimal, as in ISO-ChaI 147. The effects of the PSF overlap are minimized by reducing the aperture radius. However, we find a flux discrepancy compared to previous observations (*10*), which could be due to this overlap (see fig. S1). Estimating the background from an annulus around the source (*25*) reproduces the flux levels from the previous study, but the noise level is considerably higher preventing molecular identification and analysis. We also tried estimating the background noise from the four-point dither pattern, smoothened using a median filter. This too results in a good match with the flux levels from previous observations, but again suffers from higher residual noise. We therefore adopt the pair-wise dither subtraction. Since we use the continuum-subtracted spectrum, the flux discrepancy does not affect our analysis. The spectrum is rich with molecular features and only very small wavelength regions are line-free (based on the molecules we are able to identify). Following previous methods (*25*), we determine the noise level (1σ) between 17.1 μm-17.2 μm as 0.07 mJy.

Slab models

We generate a grid of zero-dimensional (0D) slab models, one for each of the molecules we investigated. Each model samples the temperature in steps of 25 K from 25 K to 1500 K and column densities in steps of ⅙ dex from $10^{14}$ cm$^{-2}$ to $10^{24.5}$ cm$^{-2}$. For $C_6H_6$ and $C_3H_4$ we limited the temperature to 600 K. We assume all molecules are in local thermodynamic equilibrium (LTE). 0D slab models assume a homogeneous column of gas that radiates with a fixed temperature and a constant column density. The slab models were constructed following previously described methods (*13,25*). The specific line intensity $I_\nu$ [Jy sr$^{-1}$] is obtained by

$$I_\nu = B_{\nu_0}(T)(1 - e^{-\tau_\nu}) \tag{S1}$$

where $B_{\nu_0}$ is the Planck function, $T$ is the gas temperature, $\nu$ is the frequency, $\nu_0$ is the line center frequency. We assume a Gaussian line profile. The optical depth $\tau_\nu$ at a frequency $\nu$ due to all line transitions is calculated as:

$$\tau_\nu = \sum_{i=1}^{N_{\text{lines}}} \tau_{0,i} e^{-\left(\frac{\nu - \nu_{0,i}}{\Delta \nu_D}\right)^2} \tag{S2}$$

where $\Delta \nu_D$ is the width of the line profile in Hz, $N_{\text{lines}}$ is the number of line transitions of a molecule and $\tau_{0,i}$ is the optical depth at the line center of line transition $i$ given by

$$\tau_{0,i} \equiv \frac{N A_{ul} c^2}{8 \pi \nu^2} \left(\frac{g_u}{g_l} b_l - b_u\right)$$

where $A_{ul}$ is the Einstein A coefficient of a line transition, $g$ and $b$ are the statistical weight and level population, subscripts $u$ and $l$ refer to the upper and lower energy levels, $c$ is the speed of light.

The line width is

$$\Delta \nu_D \equiv \frac{\nu_0}{c} \sqrt{\frac{2kT}{m} + v_{\text{turb}}^2} \tag{S3}$$

where $k$ is the Boltzmann constant, $m$ is the molecular mass and $v_{\text{turb}}$ is the turbulent velocity which we assume is 2 km s$^{-1}$ (*9*).

We use a fine frequency grid with a spectral resolving power of $10^5$ to calculate the model spectra as described above. Then we convolve the spectra to a spectral resolving power of 2500 (which is representative of the resolving power in channels 2 and 3 on MIRI).

We then use the PRODIMOPY (*57*) software to perform the reduced $\chi^2$ fits of the slab models to the observed spectrum, which also calculates the slab models using the above equations. We explore the parameter space of column density (*N*), temperature (*T*) and emitting area equivalent radius (*R*). The emitting radius (*R*) reflects purely geometric emitting area $\pi R^2$ and not necessarily the physical distances from the central object, i.e. the emitting area acts as a scaling factor to reproduce the absolute flux level. The column density and temperature determine the relative flux levels of lines and the shape of the continuum.

Additional molecular data

The models require molecular spectroscopic data: the line frequency, Einstein A coefficients, statistical weights, level energies, and the total internal partition sums for each molecular species. We adopted values from the high-resolution transmission molecular absorption database (HITRAN) (*58*) in most cases, with Gestion et Etude des Informations Spectroscopiques Atmosphériques database (GEISA) (*59*) for some species (Table S1). For one and two carbon-bearing hydrocarbon molecules, the isotopic fractions (single $^{13}$C-bearing isotopologues) of 70 and 35 are used in calculating the spectra (*60*). However, in some cases these databases do not contain all the values we require.

The spectroscopic data of only the fundamental excitation of only one ro-vibrational band of $^{13}$CCH$_2$ is available in the database. There are two peaks at 13.695 μm and 13.732 μm associated with $^{13}$CCH$_2$, which our models do not reproduce due to incomplete data. There are multiple smaller peaks longward of the C$_2$H$_2$ *P*-branch peaks, with lower flux level relative to the *Q*-branch peak. These might be the corresponding *P*-branch peaks of $^{13}$CCH$_2$ (fig. S2). Several features in the spectrum (fig. S2) cannot be identified with molecules for which spectroscopic data are available, such as the ~17.36 μm feature, and the region between ~15.9 μm to ~16.9 μm (Panel **C** in fig. S2).

**Line data for C$_3$H$_4$**

The C$_3$H$_4$ molecule is included in the GEISA database (*59*). For this species, the GEISA list includes only the line positions and intensities, the lower state energies, and partial information on the vibrational and rotational quantum numbers.

The vibrational transitions from a higher energy state to a ground state are called the cold band, while the transitions to a state that has higher energy than a ground state are called the hot band. In the 15.7 μm region, this GEISA list includes the contribution from the cold ν$_9$ band centered at 638.575 cm$^{-1}$ (15.66 μm) together with the two components (in ν$_9^{\pm1}$+ν$_{10}^{\pm1}$ and ν$_9^{\pm1}$+ν$_{10}^{\mp1}$, respectively) centered near 629.895 cm$^{-1}$ (15.88 μm) and 649.342 cm$^{-1}$ (15.40 μm) of the first hot band ν$_9$+ν$_{10}$-ν$_{10}$. The sum of the individual line intensities of the cold and the first hot band ($S_{\nu_9}$) and ($S_{\nu_9+\nu_{10}-\nu_{10}}$) available in the GEISA database at 296 K are:

$$S_{\nu_9} = 7.13 \times 10^{-18} \text{ cm}^{-1}/(\text{molecule. cm}^{-2}) \quad (S4)$$
$$S_{\nu_9+\nu_{10}-\nu_{10}} = 2.63 \times 10^{-18} \text{ cm}^{-1}/(\text{molecule. cm}^{-2}) \quad (S5)$$

The higher order hot bands, for which spectroscopic parameters are partially available, are not considered in the line-list.

We computed the rotational $[Z_{\text{Rot}}(T)]$ and vibrational $[Z_{\text{Vib}}(T)]$ partition functions to calculate the total partition function $[Z_{\text{Tot}}(T)]$ of C$_3$H$_4$, for the temperature range $T = 50$ K to $T = 600$ K, using the methods described by previous work [(*61*), their equations 8 and 10]. We obtain, for 296 K, the rotational, vibrational, and total partition functions as follows:

$$Z_{\text{Rot}}(296 \text{ K}) = 21275.5, \quad Z_{\text{Vib}}(296 \text{ K}) = 1.774$$
$$Z_{\text{Tot}}(296 \text{ K}) = Z_{\text{Rot}}(296 \text{ K}) \times Z_{\text{Vib}}(296 \text{ K}) = 37734.1 \quad (S6)$$

We checked the quality of the GEISA line list regarding line positions and intensities using the rotational constants of the ground states, and ν$_9$=1, ν$_{10}$=1, and the ν$_9$(=1)+ν$_{10}$(=1) vibrational states from previous publications (*62-64*). We also used integral band intensity measurements ($S'_{15.7}$) for the 15.7 μm region (*65*) to calibrate the line intensities in our line list. The integral band intensity measurement at 296 K for the 580 cm$^{-1}$ to 700 cm$^{-1}$ (14.29 μm to 17.24 μm) spectral range is:

$$S'_{15.7} = 14.96(\pm0.32) \times 10^{-18} \text{cm}^{-1}/(\text{molecule. cm}^{-2}) \quad (S7)$$

This value includes the contribution from the cold ν$_9$ band, together with those from the overall set of hot bands associated to ν$_9$. These hot bands are, the already mentioned ν$_9$+ν$_{10}$-ν$_{10}$, together with all other contributing higher order hot bands ($v_9+2v_{10}-2v_{10}$, …$v_9+nv_{10}-nv_{10}$, etc., $2v_9-v_9$, $v_9+v_8-v_8$, …, $v_9+nv_8-nv_8$, etc., where n is an integer greater than 1). We estimate the $v_9$ band intensity ($S'_{v_9}$) contribution to $S'_{15.7}$ (*65*):

$$S'_{v_9} = S'_{15.7}/Z_{\text{Vib}} 8.43 \times 10^{-18} \text{cm}^{-1}/(\text{molecule.cm}^{-2}) \tag{S7}$$

For our line list to match the measured intensities in the 15.7 µm (*65*), we multiplied all intensities by a constant factor $r$:

$$r = S'_{v_9}/S_{v_9} = 1.18 \tag{S9}$$

This list is consistent with the previously measured intensities for the 15.7 µm band of $C_3H_4$ (*68*) and includes the Einstein coefficients and the lower and upper state statistical weights. Tab. S2 provides a short description of this line list. The calculated line list is provided as data S2.

**Line data for $C_6H_6$:** Data for the benzene molecule is available in the GEISA database but includes only the line positions, intensities and the lower state energies. The existing line list involves only the cold $v_4$ band centered at 673.975 cm$^{-1}$ (14.84 µm). We therefore generated additional spectroscopic parameters.

Cold $v_4$ band: We began by determining nuclear spin statistical weights and Einstein coefficients for the existing list for the cold $v_4$ band. The Einstein coefficients were computed (*66*), using spectroscopic constants for benzene (*67*).

Partition functions: Our analysis requires the partition functions of benzene. We adopt empirical equations (*67*) to compute the vibrational and rotational contributions, $Z_{\text{Vib}}(T)$ and $Z_{\text{Rot}}(T)$ respectively, to the benzene partition function over the 50 K<$T$<500 K temperature range:

$$Z_{\text{Vib}}(T) = (1 - 7.015 \times 10^{-5}\theta + 2.396 \times 10^{-8}\theta^2 + 3.32 \times 10^{-13}\theta^3)$$
$$/(1 - 5.426 \times 10^{-5}\theta) \tag{S10}$$
$$Z_{\text{Rot}}(T) = 21.95 + 93.735 \times \theta + 7.96 \times 10^{-6}\theta^2 + 3.5 \times 10^{-10}\theta^3 \tag{S11}$$

with $\theta = T^{3/2}$.

We test how accurately these equations predict the partition functions. For a given temperature, we compute the rotational partition function by summing the contributions due to the rotational energy levels up to a given threshold energy limit. Then, we evaluated the impact of increasing this threshold limit to the computed rotational partition function. Convergence is achieved for all temperatures for calculations performed up to $J$=120; we find the computed $Z_{\text{Rot}}(T)$ value differs by 0.2% from the value computed using equations S11. Similarly, we find that computing $Z_{\text{Vib}}(T)$ using the approach prescribed in previous work [(*67*), their equation 6] differs from the value indicated by equation 11 by less than 0.5 % in the whole temperature range.

Hot bands: The hot bands contribution at 14.9 µm is not present in the GEISA line list. Benzene is a heavy molecule with 20 vibrational modes, some of which correspond to a low vibrational energy, including $v_{20}$ at 398.131 cm$^{-1}$ (25.12 µm). Therefore, at 14.9 µm, the cold $v_4$ band feature is heavily overlapped by hot bands contributions, which provide about ~45% of the flux at room temperature, according to our calculated vibrational partition function. Accounting for these hot bands requires spectroscopic parameters (band centers, rotational constants) that are not available for the upper states ($v_4+v_x$, for 20 vibrational modes x=1 to 20), and often also for the lower states ($v_x$, with x=1 to 20), involved in any (first order) $v_4+v_x-v_x$ hot band. These data are also not available for the second order ($v_4+v_x+v_y$)-($v_x+v_y$) hot bands that also contribute to the observed infrared signature.

We adopt experimental measurements (*68*) of absorption cross sections of benzene in the 7 to 15 µm region at temperatures between 235 K and 297 K. Those experiments used pure benzene and several pressure-broadened spectra (by $N_2$ and by a mixture of $H_2$ and He mixture).

We adopt the empirical pseudo line lists (*68*) derived from those experiments for benzene in $N_2$ and in ($H_2$ and He) environments that numerically describe the temperature dependence of the measured cross-sections. Each pseudo-line is listed, with a 0.005 $cm^{-1}$ periodicity, as a pseudo individual transition with an individual line position, strength, and lower state energy.

We include additional hot bands of benzene in the 14.9 μm region. In the cross sections from the pseudo line list (*68*), these hot bands appear as several *Q*-type peaks that are shifted relative to $v_4$ in the low frequency range by about ~1 $cm^{-1}$ up to ~2.2 $cm^{-1}$.

To model these contributions, we used the $v_4$ band line list as a starting point and estimated the position of several $v_4+v_x-v_x$ hot bands from the experimental cross sections (*68*). The intensity of each hot band, $S_{v_4+v_x-v_x}(T)$, relative to its $v_4$ cold band counterpart, $S_{v_4}(T)$, is assumed to be:

$$S_{v_4+v_x-v_x}(T) \approx e^{-E_x/kT} \times S_{v_4}(T) \times g \tag{S12}$$

where g is the vibrational degeneracy of the considered $v_x$ lower state (g=1 for x=1 to 10, g=2 for x=11 to 20). Because the pseudo line list provides an estimate of the temperature dependence of the investigated hot bands, we use its relative intensity to guess the (probable) vibrational assignment for the observed hot band-type *Q*- branch structures, and therefore the vibrational energy, $E_x$, for the lower state of the transition.

After estimating the band center of a given $v_4+v_x-v_x$ hot band, we use a simple frequency shift compared to the $v_4$ band to compute the line-by-line list of this hot band. The intensity of a given $v_4+v_x-v_x$ transition is deduced from its counterpart in $v_4$ that involves the same quantum numbers using equation S12. The lower state energy level is deduced from its equivalent for the $v_4$ transitions by an addition of the $E_x$ vibrational energy. As for the cold band, the Einstein coefficients were computed using a previously described method (*66*).

The contribution of hot bands centered at 671.82, 672.87, 672.90, 673.48, and 673.61 $cm^{-1}$ (or 14.885, 14.862, 14.861, 14.848, and 14.845 μm) were considered in this way. Tab. S3 gives a brief description of this data collection. The calculated line list is provided as data S3.

**Line data for $CH_3$:** Data corresponding to lines of one fundamental vibrational mode of $CH_3$ are available (*69*). Based on several hydrocarbons showing multiple emission bands and high column densities, we expect the same for $CH_3$ which can fill the missing flux in the 15.9 μm to 16.9 μm region. However, we cannot test this expectation because insufficient spectroscopic data is available.

X-ray non-detection of ISO-ChaI 147

The field of ISO-ChaI 147 was observed by X-ray Multi-Mirror – Newton telescope (XMM-Newton) (observation ID: 0203810201, principal investigator: A. Telleschi, observation date 2004-02-27, exposure time 30.7 ks). We confined our data reduction to the most sensitive European Photon Imaging Cameras (EPIC), the EPIC-pn camera (*70*), following standard pipeline procedures in the Science Analysis System (SAS, version 1.3) (*71*) and applying current calibration files (version XMM-CCF-REL-378) (*71*). High particle background levels were eliminated (*72*). Images (fig. S3) were produced in the 0.2 to 10 keV (eV is electronvolt) wide band and the narrower 0.5 to 2 keV band. The latter was chosen because the adopted spectral model with interstellar absorption (see below) showed that most of the counts are expected within this band. However, the target was undetected in both images (fig. S3).

To estimate an upper limit for the count rate and the X-ray flux, we adopt a spectral model for the X-ray source. We used the XMM-Newton Extended Survey of the Taurus Molecular Cloud (*73*) to identify detected low-mass T-Tauri stars with spectral types similar to ISO-ChaI 147 (~M5.5) and similarly moderate interstellar absorption. We select the objects 2MASS

J04150515+2808462, 2MASS J04190197+2822332, 2MASS J04190110+2819420, 2MASS J04354183+2234115, 2MASS J04265732+2606284, 2MASS J04330781+2616066, 2MASS J04315844+2543299, 2MASS J04410424+2557561. We model each source using two plasma components at temperatures of 5.5 MK and 20 MK, each with identical volume emission measures $EM = n_e n_i V$, where $n_{e,i}$ are the source electron and ion densities, respectively, and $V$ is the source volume.

The absorbing interstellar gas column density ($N_H$) is derived from the visual extinction of ISO-ChaI 147 $A_V$= 2.5 mag (*15*) using $N_H = 1.8 \times 10^{21} A_V \text{cm}^{-2}$ (*74*) to find $N_H = 4.5 \times 10^{21} \text{cm}^{-2}$. We then calculate the $EM$ that produces a flux at the detection upper limit derived from the XMM-Newton exposure.

We followed standard procedures for source detection and upper limit estimates in SAS (*72*), producing a sensitivity map (using the SAS task `esensmap`) after constructing an exposure map (`eexpmap`), performing source detection across the field (`eboxdetect`) considering an appropriate mask (`emask`), and constructing a background map using two-dimensional spline fits to point-source free regions (`esplinemap`). We set the sensitivity limit by adopting a detection likelihood $L = -\ln(p) = 6.60765$ corresponding to a probability ($p$) for a random Poissonian fluctuation to cause a detection at the (one-sided) $3\sigma$-level ($p = 0.00135$). We find this sensitivity limit to be 0.00284 ct s$^{-1}$ (or counts per second) for the 0.2 to 10 keV band and 0.00179 ct s$^{-1}$ for the 0.5 to 2 keV band.

We adjusted the two (assumed to be identical) EMs in the spectral model to the derived limiting count rates to determine upper limits to fluxes observed at Earth. After removal of the interstellar absorption and correction for the stellar distance of 191.4 pc (*14*), we determine a $3\sigma$ upper limit for the source X-ray luminosity ($L_X$) for the 0.1 to 10 keV band: $L_X < 1.42 \times 10^{29}$ erg s$^{-1}$ based on the detectability in the broad 0.2 to 10 keV detector band, and $L_X < 1.10 \times 10^{29}$ erg s$^{-1}$ based on the detectability in the 0.5 to 2 keV band. We adopt the latter value as our upper limit for the source X-ray luminosity.

We verified our analysis using the upper limit estimation software FLIX (*75*) finding very similar values ($3\sigma$ count rate upper limits of <0.023 ct s$^{-1}$ for the 0.2 to 10 keV band and <0.0017 ct s$^{-1}$ for the combined 0.5 to 1 keV and 1 to 2 keV bands). The estimated upper limit to $L_X$ is consistent with detections and upper limits for young stellar objects of similar spectral types in Taurus [(*73*), their figure 12].

Our upper limit is subject to physical uncertainties. We assumed a spectral model with two plasma components but this can vary considerably even among similar sources. The interstellar hydrogen absorption column density estimated from $A_V$ determines the spectral shape below 1 keV, and thus depends on the accuracy of the reported $A_V$ (*15*).

Modelling the JWST spectrum

Fig. 1 shows that the molecular emission in the spectrum is overlapping – no single species can be completely isolated and fitted to an individual model. Molecules that have high column densities, such as $CH_4$, $C_2H_2$, and $C_2H_4$, form wide pseudo-continua which overlap with other pseudo-continua. Emission features of molecules with lower column densities appear on top of these pseudo-continua. We therefore adopt a stepwise approach.

Photometric measurements from visible to millimeter wavelengths are required to define a dust continuum. However, we cannot define the continuum because there are no far-IR or millimeter photometry for our source and the presence of multiple molecular continua. We assume a linear continuum in the MIRI wavelength range during our model fitting, with the slope and intercept as free parameters. Molecular lines were fitted after subtracting the assumed continuum across the spectrum. Although the dust continuum is typically a combination of multiple blackbodies at different temperatures (*76*), we cannot disentangle this independently from the molecular pseudo-continua. Assuming the continuum is a straight line requires few free parameters and is computationally efficient.

A typical 10 μm silicate feature (*77*) cannot reproduce the ~10 μm region in the observed spectrum. For instance, assuming a continuum with a weak silicate feature (fig. S1) would produce continuum-subtracted flux excess longward of $C_2H_4$ and $C_2H_2$ which cannot be reproduced by molecular emission. This is because the silicate feature peaks at ~10 μm, shortward of the 10.53 μm $C_2H_4$ feature. If we assume a continuum with a silicate feature that matches the observed flux level at ~10 μm, then it overpredicts the flux at the 18 μm dust feature. We conclude that the contribution of the dust feature to the continuum is minimal and assume a linear continuum.

We identify parts of the continuum-subtracted spectrum that are dominated by emission from one molecule and fit a model of that molecule's emission in that narrow spectral range. We start with $C_2H_4$ between 8.49 μm and 11.60 μm (fig. S4). The best-fitting model of $C_2H_4$ was then subtracted, moving on to the next most dominant molecule. Regions where other molecular emissions spectrally overlap were masked from the $\chi^2$ calculation. In certain wavelength regions, two molecules almost entirely overlap; in such cases the two molecules were fitted simultaneously (thus the parameter space explored is $(N_N \times N_T \times N_R)^2$, where $N_N$, $N_T$, and $N_R$ are the number of grid points in column density, temperature, and emitting radius parameter space). Because models with large column densities produce the molecular continuum but do not reproduce the *Q*-branch peaks, we expect some radial emission gradients of molecules with very large column densities, we fit them with two models of the same species simultaneously with differing column densities. For these fits, we limit the parameter space of the column density of one model to those column densities which form a pseudo-continuum at a given temperature and the remaining parameter space to the other model. The temperatures of both models are assumed to be the same (thus the parameter space explored is $N_{N1} \times N_{N2} \times N_T \times N_R$, where $N_{N1}$ and $N_{N2}$ are the number of grid points explored in the column density space for the two models). The order of fitting is: continuum, $C_2H_4$, $C_2H_2$, $C_6H_6$, $CO_2$, ($C_3H_4+C_4H_2$), $C_2H_6$, $HC_3N$, $HCN$ and finally $CH_4$. The fitting was then repeated by redefining the continuum based on the residuals. We varied the slope of the continuum to match the flux level at 8.30 μm, where $CH_4$ and $C_2H_4$ overlap in wavelength. Then, individual molecules were fitted by subtracting the best-fitting models of all other molecules from the previous step, in the same order in the same wavelength region.

The stepwise method we use evaluates the $\chi^2$ in narrow wavelength ranges because the molecular emissions largely overlap in wavelengths. Hence, we cannot use these $\chi^2$ values to evaluate the uncertainties on the estimated properties (across the entire wavelength) in the formal way, i.e. by the 1σ or 3σ contour levels on the $\chi^2$ maps [e.g. *25*]. The stepwise approach allows a computationally feasible procedure to produce a consistent and well constrained fit. To provide some context on the errors, we changed the order of fitting the molecules and this results in changes in the derived temperatures and column densities by ±75 K and ±0.5 order of magnitude, respectively.

Due to the spectral overlap of all molecules, errors introduced by the fitting of one molecule are carried over to the next molecule in the fitting procedure, and so on. This limits our ability to provide quantitative information on molecules with weak features and upper limits for non-detections. The model fits indicate that the molecules with high column densities are spatially extended and hence our zero-dimensional models are very simple approximations. We do not account for co-spatial emission in our simple slab models, which would be important for weak emitting molecules that spectrally overlap with the molecules of high column density. Although we cannot estimate reliable upper limits on the properties of weak emitting molecules, such as HCN, $HC_3N$, and $C_2H_6$, we can still clearly detect their emission.

Subtracting the models of the molecules from the continuum subtracted spectrum shows non-zero fluxes at both shorter and longer wavelengths which can both be attributed (at least partly) to molecular emissions. At longer wavelengths there is insufficient molecular spectroscopic data available for relevant species such as $CH_3$ (see above). At shorter wavelengths, higher excitation bands of hydrocarbons are expected but do not have available spectroscopic data. For example, $C_2H_4$ data below ~6.6 μm is not available (cf. fig. S4). At high column densities, such weaker bands can contribute partly to the observed flux. However, the stellar spectral features become dominant in this wavelength range (<6.6 μm). Because the gas properties are estimated by excluding these wavelength regions, our analysis and interpretation are not affected by this unidentified emission.

The shape of the $C_2H_6$ spectrum is very sensitive to the temperature and the column density. Fig. S5 shows a fit to the spectral region of $C_2H_6$ neglecting any underlying molecular emission from $C_2H_4$ and $C_2H_2$. The fit in Fig. 1 accounts for the underlying molecular emission. The difference in these fits is possibly due to the opacity overlap of $C_2H_6$ with $C_2H_4$ and $C_2H_2$, which we have not considered in our fits. The physical distribution of the molecules in the disk emitting region largely defines the extent of this opacity interaction.

Figure S6 compares a synthetic spectrum (*78*) of a star at 3000 K and 0.04 $L_\odot$ to our observed spectrum of ISO-ChaI 147. The synthetic star spectrum matches observed spectrum below ~6 μm indicating that the flux at shorter wavelengths is dominated by the star. At wavelengths longward of ~7 μm, the flux contribution of the star drops to below 50%. We identify stellar absorption features of CO and $H_2O$ shortward of ~5.2 μm and ~6.2 μm respectively (blue and red features in fig. S6), both in the stellar model and the observed spectrum. Considering the strength of the stellar absorption features, any CO and $H_2O$ emission from the disk in these wavelengths must be weaker than the noise level. The stellar model is a typical spectrum that does not account for the specific properties of this source. Hence, we cannot completely rule out the presence of CO and $H_2O$ emission in the disk.

Though we do not detect any $H_2O$ emission from the disk, we estimate an upper limit for its column density of $<3.16\times10^{21}$ cm$^{-2}$, by assuming the same temperature and emitting area equivalent radius as $CH_4$ (the most abundant molecule detected) and using the non-detected water lines between 15.5 to 17.8 μm. This upper limit is equivalent to the number of oxygen atoms in $H_2O$ being a factor of <2.5 times that in $CO_2$, so it is possible that water is the dominant oxygen carrier. However, this upper limit is very sensitive to the assumed temperature and emitting radius because the water emission is weak (mean line strength approximately equal to the noise level).

**Supplementary Text**

Carbon enrichment processes

Hydrocarbons can be formed by breaking up of larger carbon-bearing molecules (top-down) or by combining atoms or smaller molecules (bottom-up).

Top-down route: Destruction of PAHs and carbonaceous material by UV/X-ray photons or high energy particles such as $He^+$ ions can release H atoms, C atoms, neutral and ionic hydrocarbon fragments (containing up to 5 carbon atoms) (*79,80*). While these photo-processes can produce smaller hydrocarbons in the inner disk, the hydrocarbon fragments can survive only if the environment has a high C/O ratio. No PAH features are detected in the MIRI-MRS spectrum, perhaps because the UV radiation of the central object is too weak (*15*) to excite the IR emission bands of PAHs (*81*). We cannot determine whether PAHs are present but unexcited. At high temperatures (>1000 K), carbonaceous material can undergo chemical decomposition or combustion through reactions with H, O, or OH (*82,83*). The latter processes require an initially O-rich environment to produce carbon-bearing molecules (such as $C_2H$, $CH_4$, $C_2H_2$) and oxygen-bearing molecules (such as CO, $CO_2$). The thermal sublimation temperature of carbonaceous material depends on its structure and composition. Inclusion of nitrogen, oxygen, or sulfur species [in the form of macromolecular carbon (*84*)] can lower the sublimation temperature. This process is predicted to produce a sublimation front (the soot line) at 500 K or 1200 K (*44,46*) which has been invoked to explain the carbon enrichment in the gas phase of the disk around J160532 (*13*). While macromolecular carbon sublimation could occur in ISO-ChaI 147, the composition of the gas is more diverse than the $C_2H_2$-dominated disk around J160532, which could indicate further gas processing after sublimation or different macromolecular compositions in the two sources.

Bottom-up route: Models of gas and grain surface chemistry (*85*) predict that benzene can form in the inner warm regions of disks via grain surface reactions. Experiments have shown that low temperature neutral-neutral surface reactions have no (or very small) energy barrier, including those between C and $H_2$, $O_2$, and $C_2H_2$ (*86,87*). Such reactions could lead to the formation of small hydrocarbons on cold grains in the molecular clouds well before the disk forms. Inward transport of hydrocarbon-enriched icy grains through gas drag (*30*) could lead to successive sublimation of those hydrocarbons according to their respective adsorption energies ($E_{ads}$): $E_{ads}(C_2H_6)$=2300 K, $E_{ads}(C_2H_2)$=2587 K, $E_{ads}(C_2H_4)$=3487 K, $E_{ads}(C_6H_6)$=7587 K (Table S4) (*88*), and thus to a spatial segregation of the small hydrocarbons. These molecules would quickly react further once in the gas phase, due to the short chemical and mixing timescales in the inner disk, unless the influx of icy grains from the outer disk remains high. However, the upper limits on the flux at millimeter wavelengths (*16*) imply that the outer disk pebble reservoir has already been consumed in ISO-ChaI 147. Models of inner disk chemistry predict that in environments with a C/O ratio >1 hydrocarbons are formed efficiently at high abundances (*38,40,89*) via gas phase ion-molecule and neutral-neutral reactions (*39,90*). The presence of large, saturated hydrocarbons such as $C_2H_6$ could indicate such a bottom-up gas phase chemistry route; such fragments are less abundant in the alternative top-down route.

Comparison of ISO-ChaI 147 with J160532

The central star properties of ISO-ChaI 147 and J160532 are very similar. Their stellar masses are 0.11 $M_\odot$ and 0.14 $M_\odot$ respectively, and their luminosities are 0.03 $L_\odot$ and 0.04 $L_\odot$ (*11,16*). While ISO-ChaI 147 is in the Chameleon I star forming region (median age: 1 to 2 Myr), J160532 is in the much older Upper Sco region (5 to 11 Myr). Both objects were non-detections in millimeter continuum and in CO emission line observations (*16,17,91*), setting similar upper limits on their disk dust masses. Both disks do not show the 10 μm silicate dust emission feature. However, ISO-ChaI 147 has an accretion rate (~$7\times10^{-12}$ $M_\odot yr^{-1}$) about two orders of magnitude lower than J160532 (~$8\times10^{-10}$ $M_\odot yr^{-1}$) (*11,17*). We expect higher accretion would produce more heating in the inner regions of J160532 and greater UV irradiation. The spectrum of J160532 shows emission lines of $H_2$ which are not observed in ISO-ChaI 147, consistent with this interpretation.

CO is detected in emission from the disk around J160532. In ISO-ChaI 147, we see strong stellar absorption features of CO and $H_2O$ (fig. S6). The emission of CO and $H_2O$ from the disk of ISO-ChaI 147 has an upper limit of $\lesssim$ 0.5 mJy, which is close to the noise level. This could mean that in ISO-ChaI 147 i) due to the lower heating, the inner disk excitation conditions could be unfavorable for strong CO ro-vibrational emission, ii) the inner disk structure is such that the emitting region favorable for CO emission is very small, or iii) the inner disk region is devoid of CO. $H_2O$ can emit over large temperature ranges, from hot (model temperatures $\gtrsim 200$ K) rovibrational lines (~6.26 μm) to cool (model temperatures $\lesssim 200$ K) pure rotational lines ($\gtrsim 12$ μm), but we do not detect any strong water emission in the ISO-ChaI 147 spectrum.

Although both disks are carbon rich, the dominant molecules and the diversity of carbon bearing species are different. Models of chemical networks and molecular stability studies predict abundant $CH_4$ in carbon rich environments (*41,89*). However, J160532 is dominated by $C_2H_2$. In ISO-ChaI 147, $CH_4$ is the dominant hydrocarbon. This might indicate that the inner disk of ISO-ChaI 147 (with pressures of 1 to 10 mbar) is much denser than in J160532 (pressures about two orders of magnitude lower) and thus closer to thermochemical equilibrium.

We conclude that these two objects, despite their similar stellar properties, have different physical and chemical properties in the inner disk. This might be due to the different C/O enrichment processes.

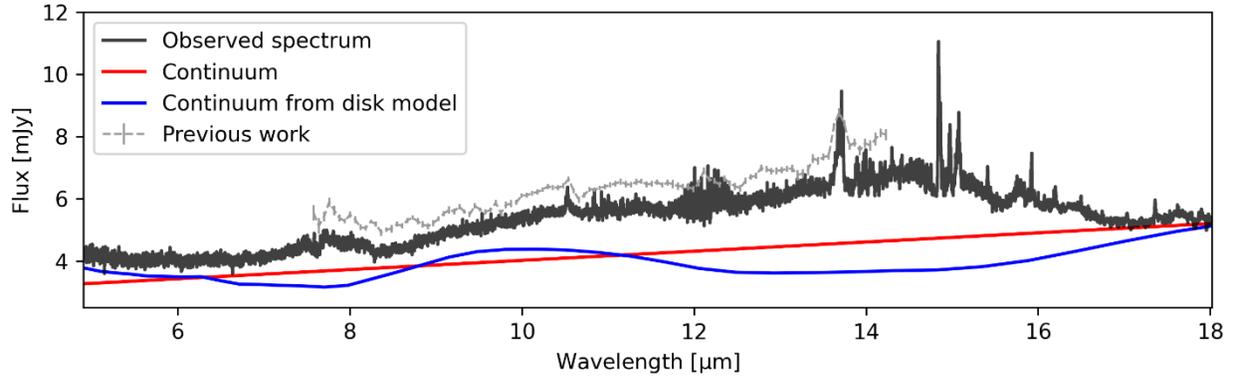

**Fig. S1. JWST/MIRI-MRS spectrum of ISO-ChaI 147.** The black line is the same spectrum as Figure 1 but before subtraction of the continuum. The gray line is a lower resolution spectrum from previous work (*10*). The red line indicates the continuum that was subtracted during the molecular model fitting process. The blue curve shows a continuum with a weak dust feature from a disk model (*26*).

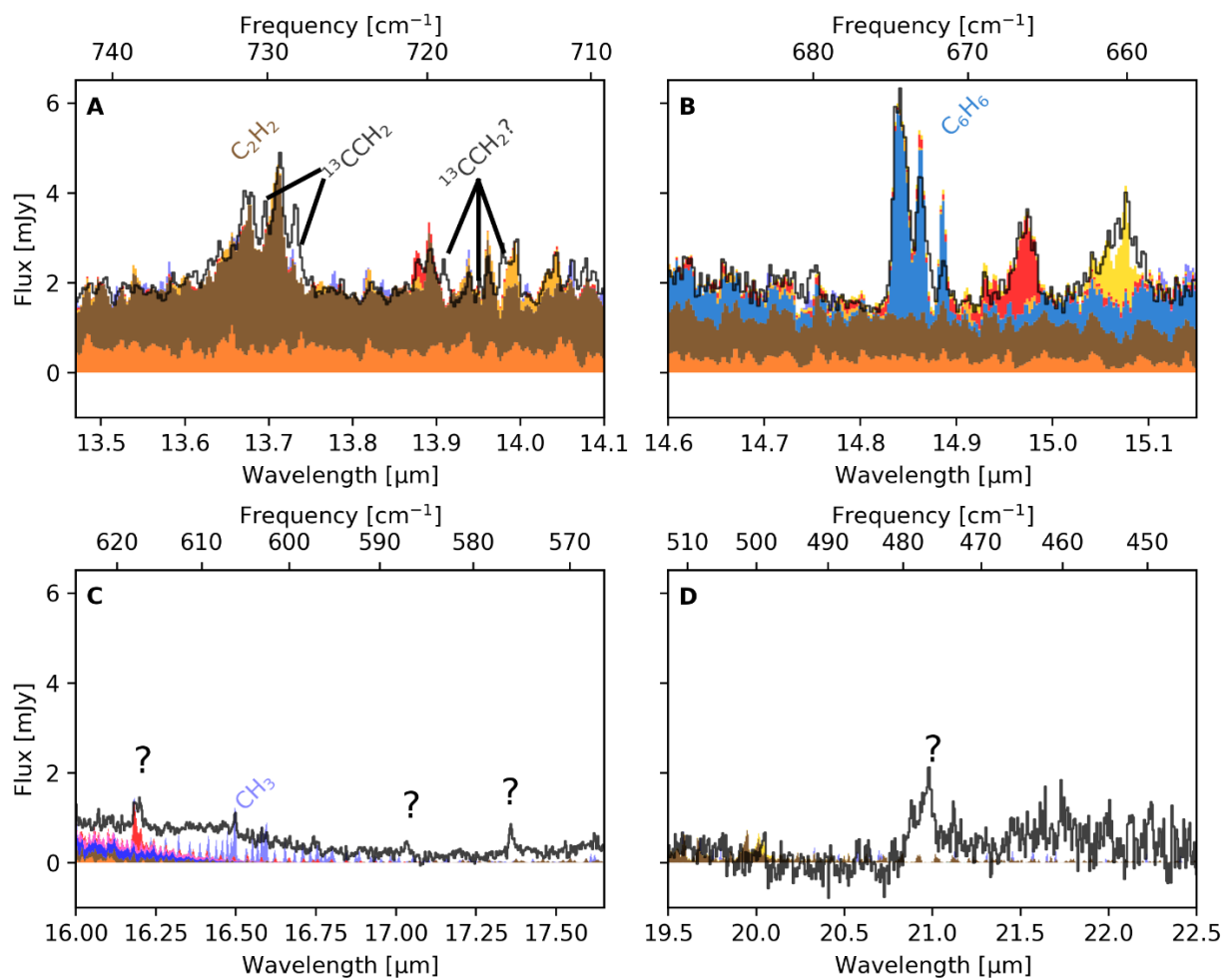

**Fig. S2. Examples of incomplete molecular data.** (**A**) potential $^{13}$CCH$_2$ features for which the molecular data is incomplete. (**B**) over- and under-prediction of fluxes of C$_6$H$_6$ at different peaks and troughs. (**C**) emission lines of CH$_3$ with incomplete molecular data and several unidentified emission features (labelled with question marks). (**D**) a strong unidentified molecular feature at 21 μm with line/continuum ratio of ~1.36. In all panels, the black line is the observed spectrum and colored shading is the same as in Fig. 1.

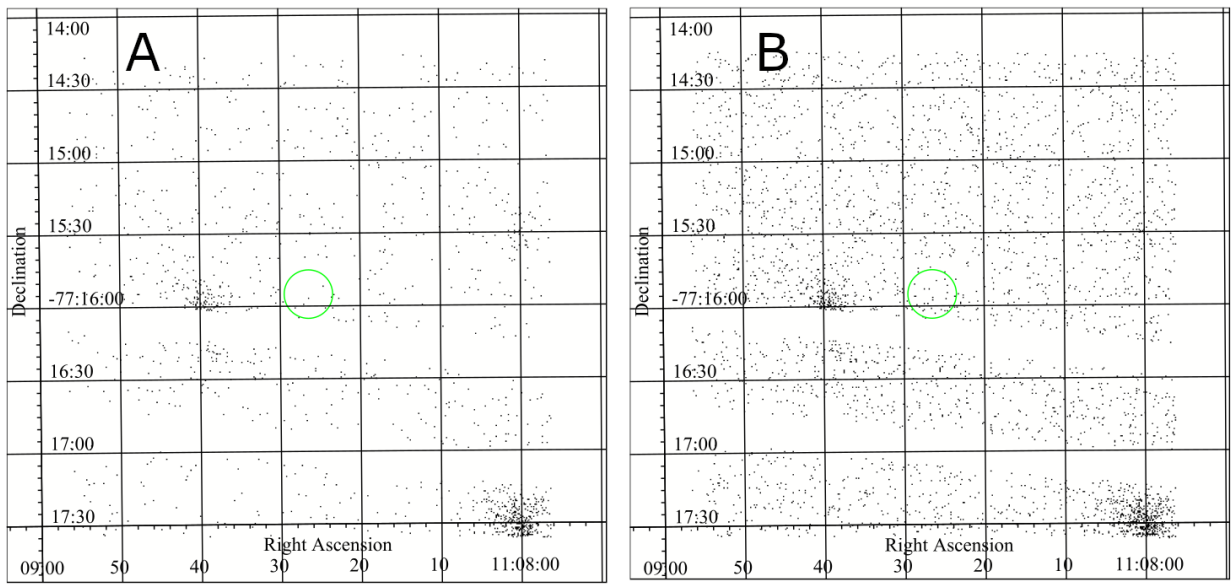

**Fig. S3. X-ray images by XMM-Newton EPIC-pn camera around the ISO-ChaI 147 position.** Panels A and B correspond to the X-ray images in the energy ranges 0.5 to 2 keV and 0.2 to 10 keV respectively. The axes are Right Ascension (RA, horizontal) in units of hours:minutes:seconds and Declination (dec, vertical) in units of degrees:arcminutes:arcseconds (1 degree = 60 arcminutes = 3600 arcseconds). The green circle is centered at the expected position of our target ISO-ChaI 147 (at RA = 11:08:26.47, dec = -77:15:55.1) but no significant source is detected. The dots distributed over the entire field are cosmic and detector background counts. The source to the east (at RA = 11:08:39.03, dec = -77:16:04.2) of the green circle is the young stellar object ISO-ChaI 151 (or 2MASS J11083905−7716042) of spectral type K5Ve, and the brighter object to the southwest (at RA = 11:07:59.95, dec = -77:17:30.6) is the young stellar object CHXR 30A (or 2MASS J11080002−7717304) of spectral type K8.

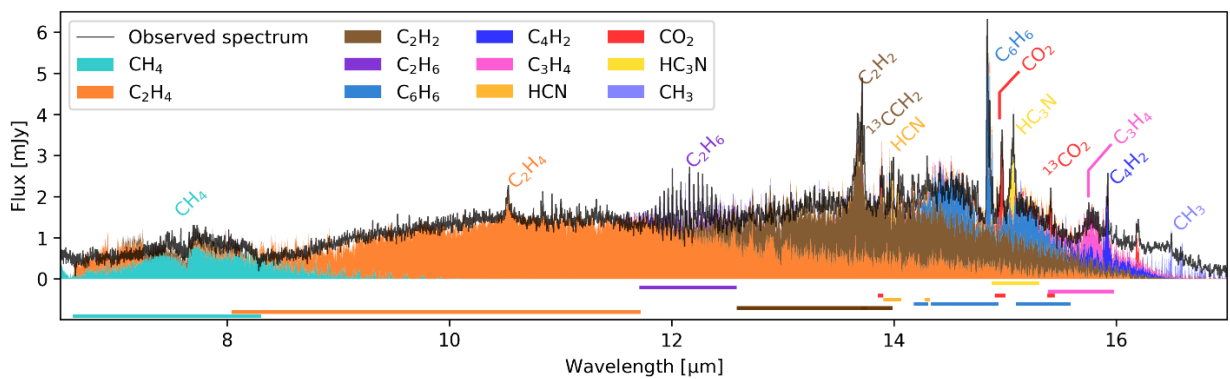

**Fig. S4. Spectral windows used for fits.** Colored regions indicate the model emission from each molecule (see legend) and the corresponding colored lines below the spectrum indicate the spectral windows used for fitting the model of each molecule. $C_3H_4$ (propyne) and $C_4H_2$ are fitted simultaneously, so they have a common spectral window.

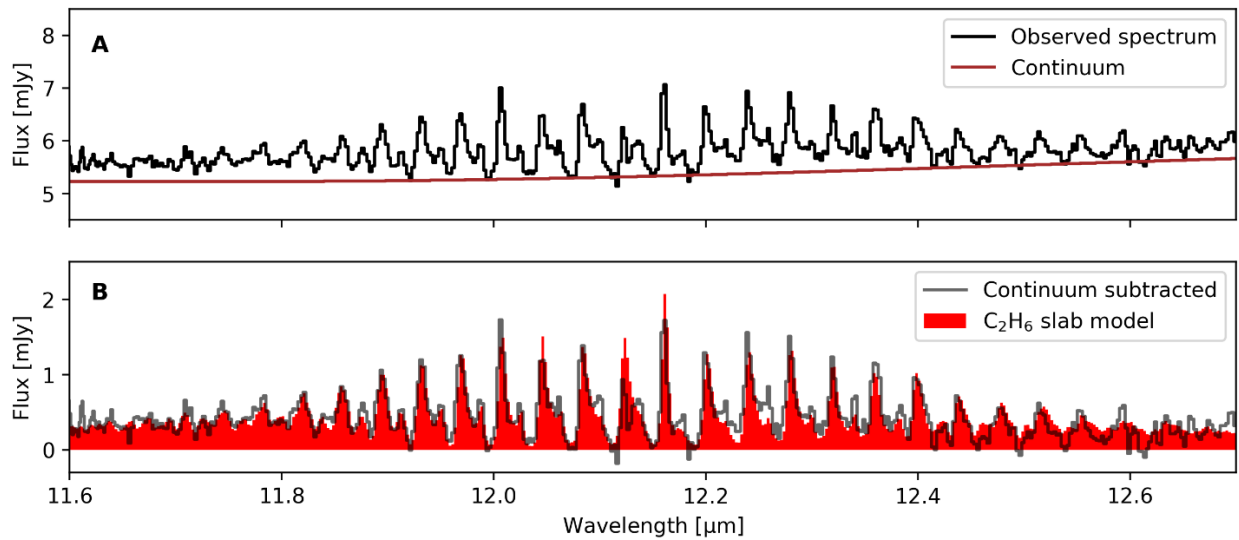

**Fig. S5. Spectrum around the lines of $C_2H_6$.** Panel A shows the observed spectrum of ISO-ChaI 147 before continuum subtraction (black) along with the smooth cubic spline (brown) used as a continuum for subtraction. Panel B shows the $C_2H_6$ emission spectrum from our model fitting (red) overlain on the continuum-subtracted MIRI spectrum (black). The $C_2H_6$ feature (spectrally) lies on top of pseudo-continua of $C_2H_4$ and $C_2H_2$. The gas temperature, column density and emitting area of the slab model spectrum are 325 K, $10^{19}$ cm$^{-2}$, and $\pi(0.07$ au$)^2$ respectively. The repeating double peak features of $C_2H_6$ appear in both the observed spectrum and the slab model.

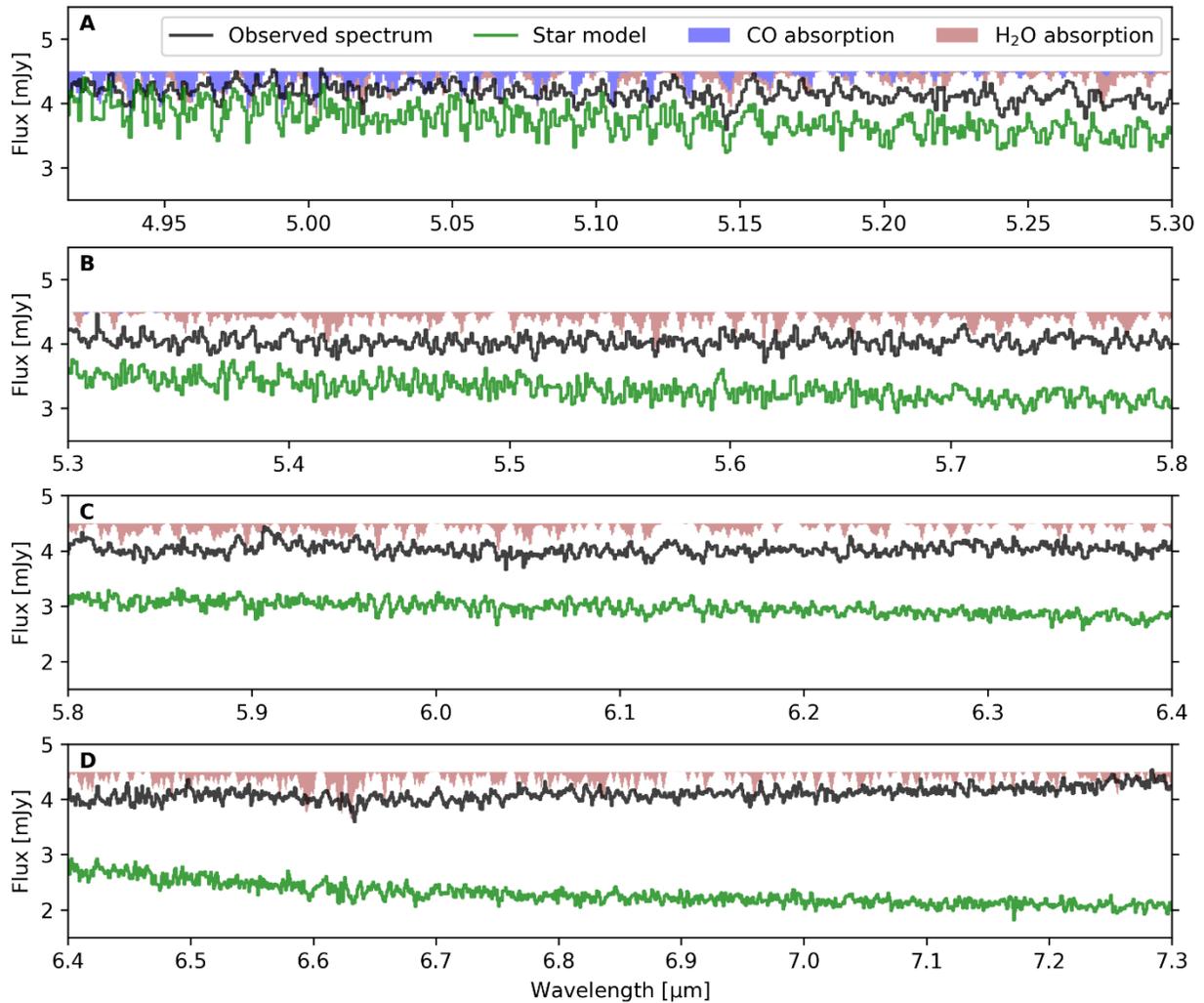

**Fig. S6. Stellar absorption features.** The observed ISO-ChaI 147 spectrum is shown in black. Synthetic spectrum of a star (*78*) is shown (green) for comparison. The blue and red colors show the absorption by CO and $H_2O$ (which have been stacked) assuming a constant 4.5 mJy continuum. Data S4 contains the spectra shown in this figure.

**Table S1.**

**Sources of molecular data used to calculate the model spectra.**

| Molecules | Source |
|---|---|
| $CH_4$, $^{13}CH_4$, $C_2H_2$, $^{13}CCH_2$, $C_2H_4$, $^{13}CCH_4$, $C_2H_6$, $^{13}CCH_6$, $C_4H_2$, $CO_2$, $^{13}CO_2$, HCN, $HC_3N$, CO | HITRAN (58) |
| $C_6H_6$, $C_3H_4$ | GEISA (59), (13), this work |
| $CH_3$ | (69) |

**Table S2.**

**Properties of the $C_3H_4$ line list.** $f_{min}$ and $f_{max}$ provide the frequency range of each band in the calculated line list.

| Band | No. of lines | Frequency (f) [cm$^{-1}$] | | Band intensity at 296 K cm$^{-1}$/(molecule.cm$^{-2}$) |
|---|---|---|---|---|
| | | $f_{min}$ | $f_{max}$ | |
| $\nu_9$ | 4103 | 592.83 | 673.48 | $8.43 \times 10^{-18}$ |
| $\nu_9 + \nu_{10} - \nu_{10}$ | 5803 | 596.18 | 672.87 | $3.11 \times 10^{-18}$ |

**Table S3.**

**Properties of the $C_6H_6$ line list.** $f_{min}$ and $f_{max}$ provide the frequency range of each band in the calculated line list. The band column refers to the probable vibrational assignment (see text).

| Band | No. of lines | Frequency (f) [cm$^{-1}$] | | Band intensity at 296 K cm$^{-1}$/(molecule.cm$^{-2}$) |
|---|---|---|---|---|
| | | $f_{min}$ | $f_{max}$ | |
| $\nu_4$ | 9797 | 642.4 | 705.3 | $8.22 \times 10^{-18}$ |
| $\nu_{20} + \nu_4 - \nu_{20}$ | 18938 | 641.3 | 704.2 | $2.37 \times 10^{-18}$ |
| $\nu_{18} + \nu_4 - \nu_{18}$ | 17560 | 642.1 | 704.8 | $8.52 \times 10^{-19}$ |
| $2\nu_4 - \nu_4$ | 8313 | 640.7 | 702.9 | $3.08 \times 10^{-19}$ |
| Total | 54608 | 640.7 | 705.3 | $1.18 \times 10^{-17}$ |

**Table S4.**

**Temperatures ($T$), column densities ($N$), and emitting area equivalent radii ($R$) of the molecules detected in ISO-ChaI 147.** Left columns are same as in Table 1 (i.e. temperature as a free parameter) and right columns assume a fixed temperature of 400 K [similar to the temperatures previously found for J160532 (*13*)]. Allowing the temperature to be a free parameter provides a better fit to the observations. The systematic uncertainties are ±75 K and ±0.5 order of magnitude in temperatures and column densities respectively, similar to Table 1. $E_{ads}$ is the adsorption energy of the species on grain surfaces (*88*). Dashes indicate species that are poorly constrained by the model fitting.

|  | $T$ as free parameter | | | $T$ = 400 K | | |
| --- | --- | --- | --- | --- | --- | --- |
| **Species** | $T$ [K] | $N$ [cm$^{-2}$] | $R$ [au] | $N$ [cm$^{-2}$] | $R$ [au] | $E_{ads}$ [K] |
| $CH_4^*$ | 300 | 1.00×10$^{23}$ | 0.048 | 2.15×10$^{20}$ | 0.029 | 1090 |
| $C_2H_4^*$ | 350 | 3.16×10$^{22}$ | 0.031 | 3.16×10$^{23}$ | 0.023 | 3487 |
| $C_2H_4$ |  | 3.16×10$^{18}$ | 0.029 | 1.00×10$^{19}$ | 0.022 |  |
| $C_2H_2^*$ | 325 | 2.15×10$^{22}$ | 0.031 | 1.47×10$^{21}$ | 0.021 | 2587 |
| $C_2H_2$ |  | 1.47×10$^{18}$ | 0.053 | 1.47×10$^{18}$ | 0.037 |  |
| $C_6H_6$ | 300 | 1.00×10$^{18}$ | 0.070 | 2.15×10$^{18}$ | 0.043 | 7587 |
| $C_3H_4$ | 250 | 1.25×10$^{18}$ | 0.056 | 5.77×10$^{18}$ | 0.025 | 2470 |
| $C_4H_2$ | 225 | 3.20×10$^{17}$ | 0.075 | 2.20×10$^{17}$ | 0.035 | 4187 |
| $CO_2$ | 225 | 2.15×10$^{19}$ | 0.088 | 6.80×10$^{17}$ | 0.043 | 2990 |
| $C_2H_6$ | - | - | - | - | - | 2300 |
| HCN | - | - | - | - | - | 2050 |
| $HC_3N$ | - | - | - | - | - | 4580 |
| $CH_3$ | - | - | - | - | - | 1175 |
| Total $\chi^2$ | 9.70 | | | 12.45 | | |

\* These species form a pseudo-continuum in the ISO-ChaI 147 spectrum.

**Data S1. (.dat file)**

**Observed spectrum and best-fitting models for each molecule.** The columns are: 1. Wavelength (μm), 2. Reduced JWST MIRI-MRS spectrum of ISO-ChaI 147, 3. continuum-subtracted spectrum, 4 to 14 are the best fitting models for 4. $CH_4+^{13}CH_4$, 5. $C_2H_2+^{13}C_2H_2$, 6. $C_2H_4+^{13}C_2H_4$, 7. $C_6H_6$, 8. $C_3H_4$, 9. $C_4H_2$, 10. $CO_2+^{13}CO_2$, 11. $C_2H_6$, 12. HCN, 13. $HC_3N$, 14. $CH_3$. Data are in Fortran notation and columns are separated by spaces. Columns 2 to 14 are in units of mJy.

**Data S2. (.dat file)**

**Calculated molecular spectroscopic data for $C_3H_4$.** The data is in the 160-byte HITRAN format (*92*).

**Data S3. (.dat file)**
**Same as Data S2 but for $C_6H_6$.**

**Data S4. (.dat file)**

**Spectra plotted in figure S6.** The columns are 1. Wavelength (microns) 2. The reduced JWST spectrum before continuum subtraction, 3. stellar model (*78*), 4. CO model and 5. $H_2O$ model. Columns 2 to 4 are in units of mJy.